# Black-Box Evasion Attacks on Data-Driven Open RAN Apps: Tailored Design and Experimental Evaluation

PRANSHAV GAJJAR*, *NextG Wireless Lab*, North Carolina State University, USA
MOLHAM KHOJA*, The University of Edinburgh, UK
ABIODUN GANIYU, *NextG Wireless Lab*, North Carolina State University, USA
MARC JUAREZ, The University of Edinburgh, UK
MAHESH K. MARINA, The University of Edinburgh, UK
ANDREW LEHANE, *6G Research R&D*, Keysight Technologies, UK
VIJAY K. SHAH, *NextG Wireless Lab*, North Carolina State University, USA

The impending adoption of Open Radio Access Network (O-RAN) is fueling innovation in the RAN towards data-driven operation. Unlike traditional RAN where the RAN data and its usage is restricted within proprietary and monolithic RAN equipment, the O-RAN architecture opens up access to RAN data (i.e., network telemetry), via RAN intelligent controllers (RICs), to third-party machine learning (ML) powered applications – rApps and xApps – to optimize RAN operations. Consequently, a major focus has been placed on leveraging RAN data to unlock greater efficiency gains. However, there is an increasing recognition that RAN data access to apps could become a source of vulnerability and be exploited by malicious actors. Motivated by this, we carry out a comprehensive investigation of data vulnerabilities on both xApps and rApps, respectively hosted in Near- and Non-real-time (RT) RIC components of O-RAN. Our investigation begins by qualitatively analyzing the O-RAN security mechanisms and limitations relevant to xApps and rApps, such as their onboarding authentication process and RIC database access mechanisms. Considering a threat model informed by this analysis, we design a viable and effective black-box evasion attack strategy targeting O-RAN RIC Apps while accounting for the stringent timing constraints (particularly for xApps) and attack effectiveness. The attack strategy employs four key techniques: the model cloning algorithm, input-specific perturbations, universal adversarial perturbations (UAPs), and targeted UAPs. This strategy targets ML models used by both xApps and rApps within the O-RAN system, aiming to degrade network performance. We experimentally validate the effectiveness of the designed evasion attack strategy and quantify the scale of performance degradation using a real-world O-RAN testbed and emulation environments. This evaluation is conducted using the *Interference Classification xApp* and the *Power Saving rApp* as representative applications for near-RT and non-RT RICs, respectively. Further, we show that the attack strategy is effective against prominent defense techniques for adversarial ML, such as defensive distillation and adversarial training.

*Equal Contribution.
Authors' Contact Information: Pranshav Gajjar*, prgajjar@ncsu.edu, *NextG Wireless Lab*, North Carolina State University, USA; Molham Khoja*, M.Khoja@sms.ed.ac.uk, The University of Edinburgh, Edinburgh, UK; Abiodun Ganiyu, aganiyu@ncsu.edu, *NextG Wireless Lab*, North Carolina State University, USA; Marc Juarez, marc.juarez@ed.ac.uk, The University of Edinburgh, Edinburgh, UK; Mahesh K. Marina, mahesh@ed.ac.uk, The University of Edinburgh, Edinburgh, UK; Andrew Lehane, andrew_lehane@keysight.com, *6G Research R&D*, Keysight Technologies, Edinburgh, UK; Vijay K. Shah, vkshah2@ncsu.edu, *NextG Wireless Lab*, North Carolina State University, USA.

# 1 Introduction

The Open Radio Access Network (O-RAN), being standardized by the O-RAN Alliance [2], represents a significant paradigm shift in mobile communications, driven by the need for cost-efficiency, diversified telecom supply chain, and innovation [6]. Traditional RAN architectures are characterized by proprietary interfaces and tightly integrated hardware and software components, leading to vendor lock-in and limited flexibility [66]. In contrast, O-RAN promotes a disaggregated architecture where components from different vendors can seamlessly interoperate through standardized, open interfaces. This modularity fosters a multi-vendor ecosystem, increased competition and accelerates the deployment of new services and technologies, thus meeting the evolving demands of mobile network operators and end-users [6, 66].

The O-RAN architecture features non-real-time (non-RT) and near-real-time (near-RT) RAN Intelligent Controllers (RICs) with corresponding third-party applications called rApps and xApps that enable data-driven RAN management and control and are key to innovation. These applications leverage various machine learning (ML) models that operate on different types of data extracted from the RAN (e.g., various key performance metrics [KPMs]) to make a wide range of critical RAN management and control decisions, such as energy saving, anomaly detection, interference detection and resource scheduling, as demonstrated by both academia and industry [12, 25, 37, 42, 67, 78]. Since ML algorithms rely on learning patterns in the collected RAN data for intelligent decision making, there is a growing concern about O-RAN applications becoming targets of potential external/internal adversaries by exploiting the data access related vulnerabilities. The openness of the O-RAN architecture reflected by open, standard O-RAN interfaces and co-located third-party applications across both the RIC frameworks expand the attack surface of the O-RAN system to include ML-based applications (i.e., xApps and rApps) as key targets [55–57]. Such attacks on applications can severely undermine their decision making accuracy, which in turn will greatly degrade network service quality, disrupting network availability, and even cause network outages.

To the best of our knowledge, except for a few survey papers [30, 41], security evaluation [7, 29, 45] and NTIA/CISA reports [1, 48] highlighting the O-RAN security issues, there is limited research that investigates the security vulnerabilities of ML based xApps and rApps against adversarial attacks, and ensuing impact on the network performance of O-RAN networks. A recent survey paper [30] reviews relevant ML use cases, analyzes the different ML workflow deployment scenarios in O-RAN, and furthermore reviews various adversarial machine learning (AML) threats, and demonstrates an AML attack on a traffic steering model. One of the best currently available resource for ML security/privacy in O-RAN are the technical reports made available by O-RAN Security Task Force Group [51]. These reports identify several ML security issues of O-RAN components (including, xApps and rApps), key threats, and their potential consequences to the network performance. Furthermore, they call for stronger authentication, authorization, and access control mechanisms to mitigate such cyber threats and privacy risks. Recently, a few research works [18, 65] have investigated training and test data control attacks for O-RAN systems (see §A.1 for discussion of related works). These early studies underscore the critical need to protect O-RAN systems and their ML-driven xApps/rApps from adversarial ML attacks while also revealing significant research challenges that we seek to address in this paper.

Motivated by the critical gaps in the literature, we conduct a thorough investigation of data vulnerabilities on both xApps and rApps, respectively hosted in Near and Non-real-time RIC components of O-RAN. First, we evaluate the security measures of O-RAN and identify their limitations, with a focus on data pipeline/management and onboarding process for xApps/rApps. Then, we explain how internal or external attackers can exploit legitimate access or system misconfigurations to obtain and alter RAN KPIs. We focus on black-box evasion attacks leveraging inference data access (via passive monitoring) that align with realistic adversary capabilities linked to the O-RAN system architecture and its security limitations. Based on this threat model, we develop an effective black-box evasion attack strategy that targets the ML models used by xApps in the Near-RT RIC and rApps in the Non-RT RIC, causing them to make erroneous decisions that ultimately degrade network performance. The proposed attack strategy consists of four key techniques: (1) *Model cloning algorithm* — that trains a surrogate model using the input inference data along with the observed hard labels/predictions made by the victim App's ML model; (2) *Input-specific perturbations* — these perturbations are generated using norm-bound and non-unbounded perturbation generation methods, out of which, the later is largely unexplored in O-RAN; (3) *Universal Adversarial Perturbation (UAP) algorithm* — that aims to generate a single perturbation vector that when added to a wide range of input data samples, causes misclassification by the target ML model. This technique aims to address the near-real-time requirements of Near-RT RIC, where input-specific perturbations may be unsuitable; and (4) *Targeted UAPs* — a specialization of UAP generation algorithm to cause the victim ML model to misclassify the perturbed input sample to a targeted class/label that is likely to cause maximum network performance degradation. The novelty of this work lies in tailoring existing black-box evasion methods, including model cloning, perturbation generation, and universal/targeted UAPs to the unique constraints of O-RAN (e.g., near-RT timing, RIC architecture). To our knowledge, this is the first study to systematically map established AML techniques into the O-RAN ecosystem and experimentally evaluate their impact using real-world testbed/emulation environments.

We evaluate the proposed black-box evasion attack strategy by considering two representative applications, one each for Near-RT RIC and Non-RT RIC, namely — (i) *Interference Classification (IC) xApp* [18] with two different implementations, KPM-based IC xApp and Spectrogram-based IC xApp, and (ii) *Power Saving rApp* [8]. We deploy and use an over-the-air LTE/5G O-RAN testbed for testing the evasion attack strategy against the IC xApp, whereas for evaluating attack effectiveness against the power saving rApp, we rely on Keysight RICTest emulator [35] to quantify the detrimental impact on network performance. Our results demonstrate that the proposed evasion attack strategy is highly effective against both applications, on par with the state-of-the-art *white-box* attack strategies [18] under all considered scenarios. Furthermore, our evasion attack strategy can be effective against prominent defense techniques for adversarial ML attacks, namely, adversarial training and distillation techniques. In summary, we make the following key contributions:

• We explore data vulnerabilities on both xApps and rApps hosted in Near and Non-RT RICs of the O-RAN architecture. We qualitatively analyze the O-RAN security mechanisms and identify the limitations relevant to xApps and rApps, such as, their onboarding process and data access mechanisms (§2.2).

• We consider black-box evasion attacks based on inference data access, aligned with the considered threat model. We discuss how internal/external adversaries can execute such attacks (§3).

• We design an effective black-box evasion attack strategy aimed at O-RAN Apps on both RIC platforms, leveraging four key techniques: model cloning, input-specific perturbations, UAPs, and targeted UAPs (§4). The designed attack strategy takes into account the O-RAN specific design considerations, such as, stringent time constraints and attack effectiveness.

• We extensively and experimentally evaluate the effectiveness of proposed attack with a prototype LTE/5G O-RAN testbed using IC xApp as the exemplar xApp (§5), and the Keysight RIC test emulator considering Power Saving as a representative rApp (§6).

• Lastly, we demonstrate the effectiveness of the proposed attack even when using prominent defense methods for adversarial ML such as adversarial training and defensive distillation (§7).

## 2 Background

### 2.1 O-RAN Architecture

The O-RAN architecture [59], illustrated in Figure 1, at its core consists of three essential functional components: the Central Unit (CU), the Distributed Unit (DU), and the Radio Unit (RU). Together, these components realize the functionality of a traditional monolithic base station and constitute the radio network segment of the mobile network architecture. Additionally, the O-RAN architecture incorporates two RAN Intelligent Controllers (RICs).

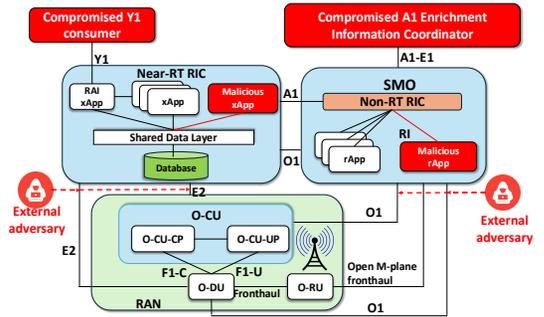

Fig. 1. O-RAN Architecture.

The near-real-time RIC (near-RT RIC) framework serves multiple AI/ML driven RAN management applications, known as *xApps*, which facilitate near-real-time control and optimization of RAN elements and resources. This function enables fine-grained data collection and execution of actions over the open, standard E2 interface, with control loops ranging from 10ms to 1s, thereby allowing for responsive and adaptive network management. In contrast, the non-real-time RIC (non-RT RIC) located within O-RAN's service management and orchestration (SMO) supports intelligent applications, known as *rApps*, which handle the control and optimization of RAN elements and resources with a granularity exceeding 1s. The rApps communicate with other RAN components through the R1 interface. The non-RT RIC also provides policy-based guidance for applications within the near-RT RIC through the A1 interface, ensuring coherent and strategic network optimization. Furthermore, it also plays a crucial role in managing AI/ML workflows, including model training and updates. O-RAN introduces a data-driven approach to RAN optimization, where real-time telemetry and network statistics (such as, uplink throughput, packet loss, spectral efficiency, PRB utilization, UE traffic patterns, SINR, RSRP etc.) are leveraged to train and refine AI/ML models underlying rApps/xApps, and to make intelligent decisions with them. These data are primarily collected via standardized interfaces — for management and performance monitoring and E2 for real-time RAN data collection and control.

### 2.2 Security Mechanisms for xApps/rApps in O-RAN Architecture and their Limitations

*2.2.1 ORAN Security Mechanisms.* In the O-RAN architecture, xApps must undergo a structured and secure onboarding process before gaining operational access within the Near-RT RIC. This ensures that only authenticated and authorized applications interact with the RIC platform and associated services. The onboarding workflow mandates the submission of the xApp with metadata and credential information to the operator. According to O-RAN Security Requirements Specification [52], during xApp registration, the xApp identifier must be explicitly associated with its authentication credentials, as outlined in *REQ-SEC-XAPP-3*[1]. The Near-RT RIC platform enforces certificate-based mutual authentication, where xApps are provisioned with operator-signed X.509 certificates. These are validated by the near-RT RIC and are required for establishing secure TLS 1.3 connections, ensuring both entity authenticity and data confidentiality (*REQ-SEC-NEAR-RT-3*

---

[1]O-RAN requirement identifiers refer to specific security mandates defined in the O-RAN WG11 Security Requirements Specification [52], issued by the O-RAN Alliance. These identifiers are used to formally track and reference compliance obligations for O-RAN components.

*and SEC-CTL-NEAR-RT-12*) requirements. The O-RAN framework also supports TLS client authentication for API access, reinforcing mutual trust and resilience against impersonation and MiTM attacks. Additionally, integrity protection of xApp packages is critical. The platform supports cryptographic integrity checks during onboarding to prevent tampering. This aligns with broader zero-trust principles recommended in the specification, including considerations for multi-factor authentication (MFA) during onboarding, where applicable.

Once successfully onboarded and authenticated, xApps may require access to platform services such as the Shared Data Layer (SDL) and RIC internal databases. The RIC platform manages such access through well-defined API interfaces and access policies. Notably, the *REQ-SEC-NEAR-RT-1* requirement mandates that the near-RT RIC must authenticate xApps before granting access to SDL during the registration phase. To ensure that xApps only interact with data they are permitted to handle, O-RAN implements Role-Based Access Control (RBAC) and supports Attribute-Based Access Control (ABAC) models. These regulate data access based on the roles assigned to an xApp or its attributes, such as function type or operational context. The combination of RBAC and ABAC offers both coarse and fine-grained control mechanisms and aligns with best practices for scalable and context-aware access governance. Furthermore, the platform enforces authorization policies using a subscription management and policy enforcement framework. This governs not only direct data access but also RAN telemetry subscriptions, which allow xApps to receive live updates while respecting privacy and resource constraints.

Similarly, rApps, which are deployed within the Service Management and Orchestration (SMO) framework and operate in the Non-RT RIC, follow a comparable security paradigm. rApps interact with the SMO and RIC components via the R1 interface and must undergo rigorous onboarding validation. According to the security specification, the SMO is responsible for authenticating rApps using X.509 certificates signed by trusted CAs *(REQ-SEC-NonRTRIC-1, SEC-CTL-SMO-5)*. API communication with the Non-RT RIC and SMO services must be secured using mutual TLS. rApps must also register their identity, roles, and operational metadata as part of the onboarding process, enabling granular authorization. Access control for rApps mirrors the xApp model. The Non-RT RIC enforces RBAC and ABAC (REQ-SEC-NonRTRIC-7 and REQ-SEC-NonRTRIC-8) to ensure that rApps access only the subset of telemetry, policy, and configuration data necessary for their authorized operations. SDL or other data repositories exposed to rApps are protected by authentication and policy enforcement mechanisms *(REQ-SEC-NonRTRIC-3)*, and integrity validation of rApp software is required before deployment.

### 2.2.2 ORAN Security Limitations.
While the O-RAN architecture introduces robust multi-layered security mechanisms to safeguard the onboarding and operation of xApps and rApps, including certificate-based authentication (*REQ-SEC-XAPP-3*), signed package validation (*SEC-CTL-NEAR-RT-1*), and fine-grained access control policies via RBAC/ABAC, the system is not impervious to compromise. Several realistic threat vectors remain even after these controls are enforced. *One major risk stems from the integrity of the xApp and rApp supply chains.* Even with the implementation of Software Bill of Materials (SBOM) and cryptographic integrity verification, malicious logic can slip in within an application. The authors in [33] demonstrate that signed xApps can exploit unguarded E2 APIs to extract data or manipulate scheduling decisions. Additionally, the NTIA Open-RAN security review acknowledges that dormant logic, outdated libraries, and backdoor payloads are often undetectable at onboarding, yet exploitable post-deployment [48]. *These risks are amplified by the reliance on shared components and third-party ML libraries.* According to [55], vulnerabilities in the AI pipeline, such as poisoned training data or compromised libraries, can impact the SDL or cause malicious decision-making in RIC optimization algorithms. *Further exacerbating this risk is the lack of runtime behavioral monitoring in most RIC implementations.* Current onboarding workflows do not incorporate continuous validation of an application's behavior post-deployment. As a result, malicious xApps or rApps can activate dormant code after successful onboarding and engage in actions such as unauthorized data exfiltration or network reconfiguration. We find similar security concerns in other domains. For example, detecting malicious mobile apps in marketplaces remains a hard problem. Despite vetting processes and advanced static and dynamic detection mechanisms, malicious apps often reach broad distribution through evading techniques like obfuscation, masking, and repackaging [14, 39]. As RIC is positioned as an open, multi-stakeholder platform, the distribution of O-RAN xApps and rApps could follow a model similar to that of mobile application marketplaces [31, 71]. Adopting such a model in the O-RAN ecosystem could introduce similar security challenges, but with potentially more severe consequences, given the potential impact that successful intrusion of malicious xApps or rApps may have on what is seen as a critical infrastructure.

*Over-permissive or misconfigured access control policies present another attack surface.* If a role granted to an xApp or rApp permits more privileges than needed, such as write access to telemetry instead of read-only, it becomes trivial for a malicious component to manipulate shared data or interfere with peer applications. Despite guidance in *REQ-SEC-NEAR-RT-1* and *REQ-SEC-NonRTRIC-8* for SDL access authentication, privilege escalation remains a credible risk in the absence of regular policy audits and fine-grained enforcement. A notable example of a misconfiguration threat is a 4G misconfiguration that affects the Virgin Media O2 UK operator, exposed device locations and types, a vulnerability caused by poor access controls [60]. In another case, a recent cybersecurity breach took place at the Japanese telecommunications company NTT Com, where hackers accessed sensitive information from nearly 18,000 organizations. The breach is believed to have resulted from unauthorized access, possibly due to compromised credentials, weak multi-factor authentication, unpatched systems, or misconfigured network devices [4, 5]. Also, a significant cybersecurity incident occurred at South Korea's SK Telecom, where a malware attack led to the leak of 26.96 million user data records. The breach was caused by inadequate protection of USIM data, resulting in government fines and mandatory security upgrades [61]. Such lapses could permit unauthorized xApp or rApp access to critical RAN data in O-RAN systems. At the Mobile World Congress (MWC) 2025 in Barcelona, the UK NCSC CTO emphasized that basic flaws like default passwords are responsible for a large proportion of telco security breaches [19], an issue similarly acknowledged by O-RAN's SFG in its warnings against poor password and credential hygiene. Finally, vulnerabilities such as hardcoded credentials, unvalidated input parsing, and insecure defaults have been identified in community-developed RIC software, such as the O-RAN SC G-release audit [74]. Though distinct from known CVEs (e.g., CVE-2023-41628 [3] affecting E2 message parsing), these findings indicate systemic weaknesses that could allow crafted xApps or rApps to destabilize or subvert the RIC control plane.

## 3 Threat Model

In this work, we consider both *internal* and *external* adversaries capable of launching black-box evasion attacks – specifically, data poisoning via inference data access – targeting ML-driven functions within the O-RAN architecture. Figure 1 depicts the position of both types of adversaries within the O-RAN architecture. Internal adversaries are malicious xApps or rApps onboarded into the Near-RT RIC and Non-RT RIC platforms respectively, while external adversaries lack direct control over RIC components but may exploit interface vulnerabilities or traffic exposure to infer sensitive information. In both cases, adversaries are assumed to have access to telemetry (KPM) data or inference outputs through legitimate or leaked API access. An example of an adversary motivated to launch such attacks is an entity linked to a nation-state actor. Such an entity could cause denial of service or service degradation by hijacking the functioning of certain key Apps (e.g., targeting the power-saving rApp and forcing certain cells to be disabled).

Importantly, the adversary's capability is focused on **inference data access** including both read and write interactions with KPMs, depending on assigned privileges. In realistic deployments, malicious xApps may gain legitimate write access to subsets of the Shared Data Layer (SDL). When access control policies (e.g., RBAC/ABAC) are misconfigured or overly permissive, an adversary may exploit this access to inject poisoned telemetry into the SDL or influence the data streams consumed by co-hosted ML models. This allows indirect manipulation of ML-driven behavior without requiring direct modification of model parameters. This aligns with recent adversarial ML literature on black-box and data poisoning attacks [18, 30, 65], and is consistent with attacker models feasible in practical RAN deployments [17, 38]

### 3.1 Internal Adversaries (Malicious xApps and rApps)

An internal adversary is an adversary who has managed to introduce a malicious xApp or rApp, possibly through supply chain compromise, misconfigured policies or certificate spoofing. These apps inherit extensive access to platform APIs, including the Shared Data Layer (SDL) and performance data repositories.

In the Near-RT RIC, a malicious xApp may exploit its legitimate access to the SDL (as per *REQ-SEC-NEAR-RT-1*) and intercept real-time KPIs and derived features used by co-hosted ML-based xApps. With the ability to monitor E2 traffic via over-permissive or unverified RBAC/ABAC policies, this xApp can extract ML inference outputs and begin learning model behavior. Using this knowledge, it can inject carefully timed perturbations (e.g., malformed KPMs) into the SDL through standard control APIs. In the Non-RT RIC, building on the misconfiguration examples in §2.2.2, nation-state actors could exploit misconfiguration vulnerabilities in more advanced ways. For example, they might create and distribute apps that appear legitimate, such as those offering power-saving features, but are designed to take advantage of O-RAN misconfigurations. A misconfigured component, like a KPI database left accessible with default credentials, could then be targeted by such apps, allowing attackers to manipulate KPI data for malicious purposes. In an even more basic scenario, a malicious rApp developed by nation-state actors could be designed to preprocess and aggregate KPIs obtained through the SMO's Performance Management (PM) system. By exploiting its role in handling KPI data, this rApp could introduce perturbations into the processed KPIs, thereby degrading the performance of other apps that depend on this data. In addition to such direct exploitation, similar threats can arise through compromised supply chains and trusted ecosystems. The authors in [36] highlight that the attacker could target a third party (such as app vendors) that provides applications to the operator and inject malicious code or modify the training dataset. Even though the apps come from trusted vendors, they may still be compromised.

Also, since rApps commonly exchange data with one another and influence higher-level ML-based policies via the A1 interface, a malicious rApp could skew inputs to downstream rApps or policy enforcers. These actions may remain undetected due to the absence of behavioral monitoring post-deployment, as we highlighted earlier in §2.2.2. Both xApps and rApps can abuse the policy registry or certificate misconfigurations to elevate privileges violating *REQ-SEC-XAPP-3* and its rApp counterparts, especially in environments with loose CI/CD validation or shared signing keys.

Both xApps and rApps share similar threat vectors: adversary gains significant autonomy, broad API access, and not subjected to post-deployment supervision. This makes internal adversaries especially credible, especially when operators adopt app-store-based onboarding models [31, 71] with third-party contributions. The credibility of these threats is further supported by findings from NTIA and WG11: vulnerabilities in CI/CD pipelines, the inclusion of outdated or unverified ML libraries, and over-reliance on static certificate checks allow conditionally activated code functions or dormant adversarial code to slip through onboarding [48, 55]. These risks mirror those identified in Android malware repackaging studies [39], reinforcing the feasibility of black-box evasion attacks even in secured RIC environments.

## 3.2 External Adversaries and Side-Channel Threats

External adversaries operate outside the RIC's logical boundary but may observe or intercept interface-level traffic to construct accurate models of ML behavior. In the Near-RT RIC, external consumers using the Y1 interface are granted access to RAN analytics via authorized APIs. While standards recommend mutual TLS-based authentication, comprehensive security evaluations for Y1 remain pending [9], unlike better-tested interfaces like A1 [68]. Malicious Y1 consumers or data-sharing partners can use live RAN state data to train surrogate models to inform physical-layer perturbations crafted to mislead real-time xApps (e.g., as in RAFA [40], where adversarial waveforms induced misclassifications in wireless classifiers without internal access). The authors in [24] has demonstrated how adversaries can exploit the Y1 interface in O-RAN to execute targeted jamming attacks without direct access to the RIC or RAN architecture. In the system model, a malicious but authenticated Y1 consumer passively subscribes to RAN Analytics Information (RAI) and forwards these metrics to an external jammer. This setup enables stealthy, analytics-driven jamming that adapts dynamically to network traffic. Their experiments show that using real-time RAN DL data, even simple threshold-based or clustering-based strategies can mimic the effect of always-on (conventional) jammer with significantly reduced energy and time overhead. This validates the Y1 interface as a high-risk exposure point, highlighting the feasibility of external analytics-based attacks against the O-RAN system. In the Non-RT RIC, similar risks arise via the A1-EI (Enrichment Information) interface, which facilitates external data ingestion into rApps. Compromised data providers, MiTM attackers on O1 links, or misconfigured APIs can enable exfiltration of inference data or facilitate adversarial feature injection.

## 4 Proposed Black-Box Evasion Attack Strategy

Evident from our threat model discussion, black-box evasion attack is a viable approach for targeting Apps on both Near-RT RIC and Non-RT RIC platforms in O-RAN. The dynamic nature of decision-making in O-RAN, coupled with the challenges in detecting adversarial perturbations, amplifies the risk of evasion attacks in both environments. In this section, we first outline the important design considerations for black-box evasion attacks targeting these RIC platforms (§4.1). We then detail the proposed evasion attack strategy that is designed using four techniques (§4.2).

### 4.1 Evasion Attack Design Considerations for O-RAN

*4.1.1 Targeted vs. General Perturbations in Evasion Attacks:* Evasion attacks can be targeted or general, depending on whether the adversary intends to force a specific misclassification or simply aims to degrade overall model accuracy.

• **Targeted Perturbation:** When an adversary knows the label space (e.g., the model classifies network conditions as *Class A* or *Class B*), they may force an incorrect decision towards a desired class. Mathematically, an adversarial input $x_{adv}$ is generated as follows:

$$x_{adv} = x + \delta, \quad \text{such that } f(x_{adv}) = y_{adv}, \quad \|\delta\| \leq \epsilon \tag{1}$$

where $x$ represents the original input features (e.g., KPM telemetry), $\delta$ is the adversarial perturbation constrained by a small budget $\epsilon$, $f(\cdot)$ is the model's decision function, and $y_{adv}$ is the incorrect target class the adversary forces the model to predict. By applying small, carefully crafted perturbations, attackers can systematically force incorrect network decisions, leading to suboptimal network performance.

• **General Perturbation:** In cases where the adversary does not know the label structure (e.g., an AI model optimizing multi-class traffic classification), the goal is to cause any form of incorrect classification without targeting a specific class. This attack is formulated as:

$$x_{adv} = x + \delta, \quad \text{such that } f(x_{adv}) \neq f(x), \quad \|\delta\| \leq \epsilon \tag{2}$$

Here, the adversary injects random perturbations into inference inputs, degrading model performance without necessarily controlling the misclassification outcome.

The general perturbations are effective against ML models (e.g., involving binary decisions), where altering one output is enough to degrade the model's performance. In contrast, the targeted perturbation will be more potent for models with multiple outputs, with the goal of forcing the model to produce a specific output. For example, disabling all capacity cells during peak hours has a greater impact than disabling only a subset of such cells, making the targeted attack more effective.

*4.1.2 O-RAN Timing Constraints and Attack Effectiveness:* The Non-RT RIC operates on inference cycles exceeding 1s (typically, within a range of several minutes or hours), enabling stealthier attacks through adversarial perturbations that may persist over longer periods. The cumulative effect of subtle perturbations over time can substantially degrade AI-driven policy optimizations, making long-term evasion attacks highly effective in Non-RT RIC environments.

In contrast, evasion attacks in the Near-RT RIC are constrained by *near-real-time execution requirements* with inference cycle ranging from 10ms to 1s. To be effective, the attack must be injected within this narrow window to influence xApp decision-making. However, due to frequent model updates and rapid Near-RT RIC control loop, the effects of the attack may be transient, necessitating sustained and precisely timed perturbations for continued impact.

## 4.2 Black-Box Evasion Attack Strategy Design

This section presents an effective black-box evasion attack strategy, leveraging established adversarial machine learning techniques. It is tailored to target both xApps and rApps on O-RAN's RIC platforms by incorporating the above outlined key design considerations and constraints. Specifically, the attack strategy targets the ML models used by xApps in the Near-RT RIC and rApps in the Non-RT RIC, causing them to make erroneous decisions that ultimately degrade network performance. The proposed strategy consists of four established techniques: (1) Model cloning algorithm, (2) Input-specific perturbations, (3) Universal Adversarial Perturbation (UAP) generation algorithm, and (4) Targeted UAPs, which together result in a very effective blackbox attack strategies against both xApps and rApps.

*4.2.1 Model Cloning Algorithm.* The evasion attack strategy leverages a Model Cloning Algorithm (MCA) to train a surrogate model using input inference data along with the observed hard labels or predictions made by the victim ML model (corresponding to a specific xApp or rApp in Near-RT RIC or Non-RT RIC, respectively). This approach satisfies the key design requirement that the evasion attack must be black-box, meaning it does not require knowledge of victim ML model's architecture, and instead relies solely on observable inputs and outputs.

The MCA algorithm is designed with two primary objectives. First, it seeks to replicate the predictions of the victim ML model as accurately as possible. A higher degree of similarity between the predictions of the victim and the surrogate model enhances the effectiveness of the *black-box* evasion attack strategy [58]. Second, the MCA algorithm is optimized to efficiently detect model convergence, thus avoiding redundant computational costs. During training, the validation metrics of an ML model typically plateau after a certain number of iterations. Beyond this point, only marginal changes occur in the metrics, leading to unnecessary computational expenses. The MCA algorithm detects this convergence and halts training early, improving computational efficiency.

While our MCA draws on established surrogate model training techniques from the adversarial ML literature, our novelty lies in tailoring these steps to the unique constraints of the O-RAN ecosystem. Furthermore, we demonstrate the MCA's effectiveness through rigorous evaluation on a real-world O-RAN testbed, highlighting its practical viability in this domain.

The pseudocode for the MCA algorithm is presented in Algorithm 1 (see §A.2). The MCA algorithm has the following key five steps. (I) First, we generate a dataset $\mathcal{D}_{clone}$, which is used to train a surrogate model. (II) Second, we perform a stratified train-test split to avoid class imbalance issues by ensuring that each subset has the same proportion of class labels. (III) Third, we select a list of classifier architectures $\mathcal{A}$ and utilize **Early Stopping**[2] and a **Learning Rate Scheduler**[3] [69] for every model in $\mathcal{A}$. (IV) Fourth, we select the model $\mathcal{M}_c$ from the list $\mathcal{A}$ that achieves the highest validation accuracy (also, refered to as cloning accuracy) and (V) Finally, the model $\mathcal{M}_c$ is returned as the surrogate model.

*4.2.2 Input-Specific Perturbation.* Once the surrogate model, $\mathcal{M}_c$, with the highest validation accuracy is obtained, our evasion attack strategy proposes to employ state-of-the-art input-specific perturbations. These perturbations are generated using Perturbation Generation Methods (PGMs), which are typically classified into two categories: norm-bounded and norm-unbounded attacks [34]. Notably, norm-unbounded attacks remain unexplored in the context of O-RAN and adversarial attacks. We consider representative methods from both categories: FGSM [26] and PGD [43] for norm-bounded attacks, and C&W [15] with DeepFool [47] for norm-unbounded attacks. These approaches are selected for their proven effectiveness in generating adversarial perturbations while balancing stealth and computational efficiency. Detailed mathematical formulations and implementation specifics are provided in Section A.3.

*4.2.3 UAP Generation Algorithm.* Input-specific perturbations typically require significant computational resources and we empirically show that in §5.3.3. While this overhead may be acceptable for the Non-RT RIC, which is usually hosted on remote servers with sufficient compute and memory capabilities, and where rApps operate on non-real-time schedules (ranging from several minutes to hours), it is impractical for Near-RT RIC xApps. As xApps have near-real-time requirements, input-specific perturbations are unsuitable. To address this challenge, we utilize UAPs. UAPs generate a single perturbation that can be applied to any input data, significantly reducing computational overhead and ensuring compliance with the Near-RT RIC's stringent sub-1s closed-loop latency requirements. *The primary goal of UAPs is to create a single perturbation vector that, when added to a wide range of input samples, causes misclassification by the target classifier.* As in [46], the UAP can be defined as follows. Let $\mathcal{S}$ be a distribution of input data, and let $C$ denote a classifier that outputs a label $C(\mathbf{x})$ for each input sample $\mathbf{x}$. Our objective is to find a perturbation vector $\mathbf{u}$ such that the classifier misclassifies most of the inputs from $\mathcal{S}$ when perturbed by $\mathbf{u}$. Formally, we aim to find a vector $\mathbf{u}$ such that:

$$C(\mathbf{x} + \mathbf{u}) \neq C(\mathbf{x}) \quad \text{for most} \quad \mathbf{x} \sim \mathcal{S} \tag{3}$$

To achieve this, we seek a perturbation $\mathbf{u}$ that satisfies the following constraints:

1. $\|\mathbf{u}\|_p \leq \epsilon$ and 2. $\Pr_{\mathbf{x} \sim \mathcal{S}} (C(\mathbf{x} + \mathbf{u}) \neq C(\mathbf{x})) \geq 1 - \zeta$,

where $\epsilon$ controls the magnitude of the perturbation vector $\mathbf{u}$, and $\zeta$ quantifies the desired fooling rate for the input data sampled from the distribution $\mathcal{S}$ [46].

The algorithm is designed to compute UAPs proceeds iteratively over the dataset and gradually constructs the universal perturbation vector $\mathbf{u}$. At each iteration, the algorithm computes the minimal perturbation $\Delta \mathbf{u}_i$ that sends the current perturbed sample data $\mathbf{x}_i + \mathbf{u}$ to the decision boundary of the classifier. This perturbation is then aggregated to the current instance of the universal perturbation. If the current universal perturbation $\mathbf{u}$ does not fool a data point $\mathbf{x}_i$, we solve the following optimization problem to find the minimal perturbation $\Delta \mathbf{u}_i$:

$$\Delta \mathbf{u}_i \leftarrow \arg \min_{\mathbf{r}} \|\mathbf{r}\|_2 \quad \text{s.t.} \quad C(\mathbf{x}_i + \mathbf{u} + \mathbf{r}) \neq C(\mathbf{x}_i) \tag{4}$$

---

[2]The validation loss $L_{\text{val}}(t)$ at each epoch $t$ is monitored, and if its does not improve for a specified number of epochs, known as patience parameter $k$, the training process is halted.
[3]The learning rate $\eta$ is reduced by a factor $\gamma$ if the validation loss does not improve for a specified number of epochs $m$

| Victim: SOTA | eps = 0.05 | | eps = 0.1 | | eps = 0.2 | | eps = 0.3 | | eps = 0.5 | |
|---|---|---|---|---|---|---|---|---|---|---|
| | Accuracy | APD | Accuracy | APD | Accuracy | APD | Accuracy | APD | Accuracy | APD |
| Base + FGSM | 0.548 | 0.048 | 0.220 | 0.097 | 0.058 | 0.188 | 0.030 | 0.288 | 0.017 | 0.400 |
| Base + UAP (FGSM) | 0.948 | 0.015 | 0.902 | 0.030 | 0.729 | 0.060 | 0.603 | 0.091 | 0.358 | 0.151 |
| DenseNet + FGSM | 0.918 | 0.050 | 0.872 | 0.100 | 0.662 | 0.194 | 0.398 | 0.277 | 0.221 | 0.413 |
| **DenseNet + UAP (FGSM)** | 0.994 | 0.010 | 0.988 | 0.016 | 0.892 | 0.064 | 0.745 | 0.096 | **0.373** | 0.159 |
| MobileNet + FGSM | 0.965 | 0.050 | 0.899 | 0.100 | 0.636 | 0.194 | 0.489 | 0.278 | 0.382 | 0.414 |
| MobileNet + UAP (FGSM) | 0.993 | 0.016 | 0.969 | 0.032 | 0.927 | 0.063 | 0.847 | 0.095 | 0.598 | 0.159 |
| 1L + FGSM | 0.983 | 0.050 | 0.972 | 0.100 | 0.810 | 0.194 | 0.568 | 0.281 | 0.472 | 0.425 |
| 1L + UAP (FGSM) | 0.997 | 0.012 | 0.992 | 0.025 | 0.974 | 0.049 | 0.898 | 0.074 | 0.720 | 0.122 |
| ResNet + FGSM | 0.790 | 0.050 | 0.707 | 0.100 | 0.503 | 0.194 | 0.364 | 0.278 | 0.258 | 0.414 |
| ResNet + UAP (FGSM) | 0.993 | 0.017 | 0.963 | 0.034 | 0.933 | 0.067 | 0.895 | 0.101 | 0.764 | 0.168 |

Table 1. Results of the primary experiment which leverages the SOTA model [18] as the victim xApp and the aforementioned models as model cloning strategies.

The perturbation vector $\mathbf{u}$ is then updated and projected onto the $\ell_p$ ball of radius $\epsilon$ centered at zero to ensure the norm constraint is satisfied:

$$\mathbf{u} \leftarrow \mathcal{P}_{p,\epsilon}(\mathbf{u} + \Delta \mathbf{u}_i) \tag{5}$$

where $\mathcal{P}_{p,\epsilon}$ denotes the projection operator. The detailed algorithm is provided in Algorithm 2 (see §A.2). It is also important to note that as UAPs are modeled as an optimization problem, it is possible to leverage different perturbation generation methods (PGMs) in §A.3 to generate perturbation masks (i.e., for computing $\Delta \mathbf{u}_i$) [49].

*4.2.4 Targeted UAPs (in short, TUP).* In xApps/rApps with binary classification tasks, a general UAP algorithm (presented in §4.2.3) is sufficient, as the primary objective is to flip the predicted class regardless of the target. However, in multi-class classification problems commonly encountered in O-RAN applications, targeted UAPs may become more strategic. By deliberately perturbing inputs to mimic a specific target class, attackers can induce systematic misclassifications that maximize operational degradation.

So, as the final technique, our evasion attack strategy adapts the UAP generation algorithm to a targeted setting by modifying the optimization constraint in Equation (4) to enforce misclassification towards a predefined target class $t$:

$$\Delta \mathbf{u}_i \leftarrow \arg \min_{\mathbf{r}} \|\mathbf{r}\|_2 \quad \text{s.t.} \quad C(\mathbf{x}_i + \mathbf{u} + \mathbf{r}) = t \tag{6}$$

This ensures perturbations push samples across decision boundaries specifically towards $t$, which we select based on its potential to maximally disrupt network performance. For example, in multi-class xApps and rApps (as demonstrated with power saving rApp in §6), we choose $t$ as the class associated with the most conservative resource allocation policy, thereby inducing persistent over-provisioning or under-utilization. The projection step in Equation (5) remains unchanged to maintain stealthiness. This targeted approach enhances attack impact in multi-class settings while preserving the computational efficiency of universal perturbations.

## 5 Evaluation of the Proposed Attack Strategy against xApps over Near-RT RIC

In this section, we assess the proposed black-box evasion attack strategy in the Near-RT RIC setting with a representative *Interference Classification (IC) xApp*. This IC xApp was utilized in a recent work [18] for white-box attacks in the O-RAN context. We prototype a LTE/5G O-RAN testbed similar to that in [18] for evaluation purposes (See §A.4).

## 5.1  Interference Classification (IC) xApp

*Interference Classification (IC) xApp* aims to detects uplink interference signals from jammers in cellular networks [18] [23] [65]. Note that this IC xApp has been leveraged in prior work to prototype white-box data poisoning attacks in O-RAN [18], with two different versions, namely: *Spectrogram-based IC xApp* and *KPM-based IC xApp*. We evaluate the proposed black-box evasion attack strategy with both these implementations of IC xApp as representative xApps in our work.

The Spectrogram-based IC xApp is based on a CNN model with four 2D Convolutional layers with a filter size of (3,3), ReLU activation [50], and filter sizes of [16, 16, 32, 32], respectively. This is followed by a Dense layer with 32 Neurons and a SoftMax [50] classification layer [18]. In contrast, the KPM-based IC xApp uses a DNN model with three ReLU-activated Dense layers with [64, 32, 16] neurons and a SoftMax classification layer. We refer to these as **victim xApps** henceforth. Since both the implementations of IC xApp are primarily binary ML classification tasks (interference or not), our black-box attack strategy employs a generalized UAP algorithm (instead of TUP).

For detailed information on the prototyped LTE/5G O-RAN testbed, refer to §A.4, and for the spectrogram/KPM dataset generation process, see §A.5. These resources are used to evaluate the proposed black-box evasion attack strategy against both IC xApps in the Near-RT RIC setting.

## 5.2  Evaluation Metrics

*5.2.1  ML Metrics.* We consider three metrics: **Accuracy**, **F1-Score** [76], and **Average Perturbation Distance (APD)**. The *Accuracy* is defined as the ratio of correctly predicted instances to the total instances, providing a measure of the overall effectiveness of the model in correctly classifying the given inputs. The *F1 score* is the harmonic mean of precision and recall, providing a balance between false positives and false negatives under adversarial conditions [76]. *APD* provides insight into the perceptibility of the perturbations and helps in evaluating the trade-off between attack effectiveness and perturbation visibility. It can be calculated as follows:

$$\text{APD} = \frac{1}{N} \sum_{i=1}^{N} \|\mathbf{x}'_i - \mathbf{x}_i\|_2 \tag{7}$$

where $\mathbf{x}'_i$ is the adversarial sample, $\mathbf{x}_i$ is the original sample and $N$ is the total number of perturbed samples. To gradually enhance the potency of the attack or the perturbation budget we use the concept of epsilon ($\epsilon$) which indicates the magnitude of the perturbation applied. This also helps us see the trade-off between the $\epsilon$ value and the perceptibility of perturbations through APD values.

*5.2.2  Network Performance Metrics.* The key network metrics used for evaluations are **throughput**, **block error rate (BLER)**, and **modulation and coding scheme (MCS)**.

*Throughput*, typically measured in megabits per second (Mbps), indicates the rate of successful data transmission over the network, reflecting its efficiency and capacity. *Block Error Rate (BLER)* is the ratio of the erroneous data blocks to the total transmitted blocks, useful for accessing the communication link quality. *Modulation and Coding Scheme (MCS)* defines the modulation type and coding rate used for data transmission; it determines the trade-off between data rate and network performance against errors.

## 5.3  Results

We first evaluate the proposed black-box attack against baseline methods using ML metrics, then assess their impacts on network performance. Unless stated otherwise, we use the spectrogram-based IC xApp and its CNN model (see §5.1) as the victim xApp and base model, respectively.

*5.3.1  Comparative analysis for different surrogate models.* Table 1 presents the accuracy and APD results for various surrogate architectures—DenseNet, MobileNet, ResNet, and 1L—assessed during Step 1 (Model Cloning Algorithms) with perturbed spectrograms at different $\epsilon$ levels. These architectures were selected as they span state-of-the-art vision models with proven robustness [16, 32], lightweight yet efficient designs [64], and a minimal single-layer baseline (1L), thereby covering a respective spectrum of computational and accuracy trade-offs. In this evaluation, Step 2 (UAP/perturbation generation algorithm) is fixed to FGSM. The base model corresponds to the CNN architecture used in the considered Spectrogram-based IC xApp, which represents the best-case scenario for the model cloning algorithm. The cloning accuracies for the Base, DenseNet121, ResNet50, MobileNetV2, and 1L models are 0.989, 0.996, 0.988, 0.828, and 0.623, respectively (under no attack, i.e., $\epsilon = 0$, not shown in Table 1). These values indicate the extent to which each surrogate model successfully replicates the base model's output predictions.

The use of input-specific perturbations, as expected, is more potent than UAPs, with an average accuracy decrease of 0.28 for the considered base model at a particular eps value; however, required a 42.063% higher APD for conducting the attack, which substantially affects the perceptibility. When we consider the cases of standard attacks and equivalent UAPs for similar APD values, the UAPs provide superior results for all the cloned architectures. *This is evident when we compare the average APD for the UAP-based attacks at eps = 0.5, which is 0.152, and the average APD for input-specific attacks at eps = 0.2, which is 0.194. Despite having a 24.278% lower APD, the UAPs on average showcased a 3.9% bigger drop in the victim xApp's accuracy.* The inferior performance of input-specific perturbations against UAPs is further seen in §5.3.3 and §5.3.6 which prove the incompatibility of such attacks in the near-RT RIC due to a substantially higher computational requirement. Among the evaluated configurations for UAP, the DenseNet emerged as the most effective surrogate architecture. The DenseNet + UAP configuration achieved performance results closely matching the best-case scenario represented by the Base model.

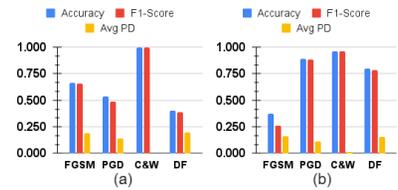

At $\epsilon = 0.5$, the DenseNet + UAP achieved an accuracy of 0.373, which is only 0.015% higher than the Base + UAP (the best-case black-box attack). This is followed by MobileNet, with the average accuracy difference of the victim xApp between DenseNet + UAP and MobileNet + UAP being 8.567% across the different epsilon values. Furthermore, the 1L + UAP configuration demonstrated significant drops in accuracy, highlighting that even simpler architectures like 1L can effectively exploit UAP to degrade the victim xApp's performance, and can even outperform popular deeper convolutional architectures. As the DenseNet + UAP provides the best result, the remaining experiments are conducted with DenseNet121 as the surrogate model. We also use a reference line, which is "Accuracy = 0.5" to assess efficacy; as a 50% accurate model can be considered as a poor classifier and equivalent to a random predictor. This helps us understand the epsilon requirement and related APD values for performing a successful attack.

Fig. 2. Black-box attacks with different PGMs—(a) input-specific perturbation, (b) UAP. DF denotes DeepFool; surrogate = DenseNet.

### 5.3.2 Analysis of PGMs.
To evaluate different PGMs, we use an eps ($\epsilon$) of 0.5 for UAPs and an eps of 0.2 for the standard attacks to understand their performance with an equivalent APD. We use a total of 350 observed predictions of the victim xApp to generate the perturbations; the same spectrograms are used across the four PGMs (i.e., FGSM, PGD, C&W, and DeepFool) to avoid biases. From Figure 2 , we can infer that for an input-specific perturbation-based attack approach, DeepFool is the best PGM method for generating perturbations; however, in the case of UAPs, FGSM provides the best results. UAPs in general show a better performance than the input-specific perturbation, and FGSM even performs better than PGD, which can be attributed to the iterative and random initialization of UAPs.

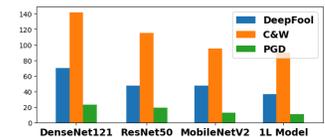

Fig. 3. Average time to generate a single perturbation (y-axis in seconds).

*5.3.3 Timing Evaluation.* From §4.2.2 and §A.3, we can say that every PGM method (except FGSM) is iterative and would leverage multiple gradient computations for a specific input data, leading to a slower attack. Even computing a single perturbation with FGSM for deeper surrogate models, MobileNetV2 and DenseNet121, it respectively takes 1.4058 and 4 seconds to compute a single perturbation, which exceeds near-RT RIC's latency threshold (< 1s). Given the significant delay in perturbation generation, MobileNetV2 will result in 64.5% spectrograms being unperturbed, while DenseNet121 misses 87.5% of spectrograms, significantly affecting its practicality for real-time O-RAN applications. To understand the performance of the other perturbation generation algorithms from §4.2.2, we compute the average prediction times for fifty $128 \times 128$ spectrograms for the tested classification architectures from $\mathcal{A}$, and the obtained results are shown in Figure 3[4]. It is clear that generating perturbations using iterative PGM algorithms (i.e., DeepFool, C&W and PGD) is highly time-consuming and does not meet the near-RT RIC's latency constraints. Norm-unbounded methods are notably slower compared to their norm-bounded counterparts, with the C&W algorithm being the most computationally intensive[5].

*5.3.4 White-Box vs. Black-Box Attacks.* Here, we compare the performance of the proposed black-box attack strategy, against the white-box attack [18]. The results, shown in Figure 4 (a), indicate that the black-box attack achieves comparable performance, with only a 0.09 epsilon difference. This shows the efficacy of the proposed attack strategy.

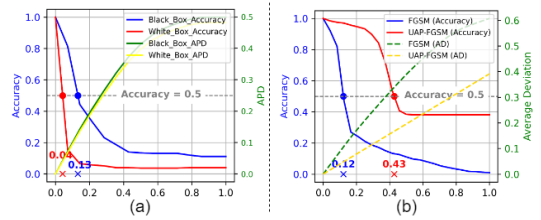

Fig. 4. (a) White- vs. black-box attack and (b) black-box attack on KPM-based xApp (x-axis: $\epsilon$).

*5.3.5 Experiment with KPMs.* The Figure 4 (b) showcases the experiment in a black-box attack scenario for the other KPM-based IC xApp[6]. The UAP is generated by the same observed dataset of the victim KPM-based xApp (See §A.5). Similar to the primary experiment in §5.3.1, where the base attack is more potent at a particular eps value, it has a substantially higher APD and the UAP can have a successful attack with a substantially small AD value.

*5.3.6 Network Performance Results.* From §5.3.3, we note that DenseNet121 fails to process 87.5% of spectrograms, making it ineffective for O-RAN's Near-RT RIC scenarios. Its slow execution time results in minimal impact on the victim xApp, allowing it to function correctly most of the time. Consequently, we select MobileNetV2, the second-best performing model, as our surrogate model. We then employ our UAP algorithm with FGSM and compare its effectiveness against a baseline FGSM with input-specific perturbations, evaluating the network performance in both cases. For each scenario, the network was set up with the UE connected to the RAN, and uplink traffic was generated using iperf3. A jammer was then introduced to evaluate the system's performance under different black-box attack strategies.

---

[4]We do not include the FGSM results in this graph due to the huge difference between the computation time of FGSM and other iterative attacks.

[5]We also evaluated the time taken and the utilized resources for training the surrogate model, and usual trends for convergence were directly linked to the model size, with the 1L model being the most resource and time efficient and the ResNet50 being the slowest to converge with the highest resource requirement.

[6]The trained ML model for the victim KPM-based xApp has an accuracy of 0.979, and the surrogate model has a cloning accuracy of 0.977.

| Models | $\epsilon = 0.05$ | | | $\epsilon = 0.1$ | | | $\epsilon = 0.2$ | | | $\epsilon = 0.3$ | | | $\epsilon = 0.5$ | | |
|---|---|---|---|---|---|---|---|---|---|---|---|---|---|---|---|
| | TASR | NTASR | APD | TASR | NTASR | APD | TASR | NTASR | APD | TASR | NTASR | APD | TASR | NTASR | APD |
| Base | 8.771 | 15.403 | 0.223 | 18.748 | 33.137 | 0.409 | 41.225 | 66.376 | 0.774 | 52.408 | 79.154 | 0.907 | 78.639 | 95.286 | 1.386 |
| MobileNet | 1.249 | 7.468 | 0.214 | 14.555 | 25.157 | 0.465 | 11.385 | 34.596 | 0.671 | 20.868 | 49.781 | 1.005 | 68.821 | 90.010 | 1.157 |
| ResNet | 3.313 | 6.056 | 0.151 | 1.404 | 16.077 | 0.382 | 37.641 | 58.471 | 0.758 | 30.220 | 59.051 | 0.861 | 59.138 | 89.606 | 1.490 |
| DenseNet | 5.983 | 12.204 | 0.232 | 1.738 | 16.588 | 0.391 | 0.0 | 23.233 | 0.775 | 44.613 | 69.677 | 1.044 | 32.352 | 63.771 | 1.450 |
| 1L | 5.866 | 8.905 | 0.227 | 14.504 | 25.096 | 0.431 | 18.617 | 55.954 | 0.727 | 19.978 | 57.648 | 0.964 | 32.668 | 68.484 | 1.697 |

Table 2. TASR, NTASR, and APD values for different models at various $\epsilon$ levels

The cumulative distribution function (CDF) plot for MCS, throughput, and BLER for various black-box attack strategies are shown in Figs. 5a, b, and c, respectively. Under no attack, the victim xApp accurately detects inter-ference and sends control to the RAN to use adaptive MCS, dynamically adjusting to maintain low error rates and optimal throughput. However, during proposed black-box attack strategy (with UAPs), the xApp fails to detect interference correctly, causing the RAN to use a fixed MCS, leading to significant performance degradation, higher BLER, and lower throughput. The MobileNet-based input-specific perturbation attack performs poorly due to its slower execution, which occasionally allows the xApp to make correct predictions. This variability results in inconsistent impacts on network performance.

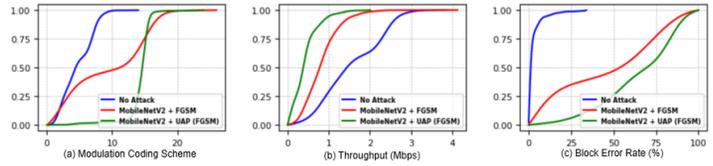

Fig. 5. CDF of various network performance metrics under no attack and the two black-box adversarial attack strategies.

## 6 Evaluation of the Proposed Attack Strategy against rApps over Non-RT RIC

Here, we evaluate the proposed evasion attack strategy for Non-RT RIC using a representative *Power Saving rApp*. By introducing subtle adversarial perturbations to the input inference data, we demonstrate how an adversary can manipulate the rApp into making suboptimal power-saving decisions. These misled decisions can lead to unnecessary cell deactivations, ultimately degrading network performance. Considering the large geographical scope of rApp decisions spanning multiple cells, we perform the evaluations leveraging the Keysight RICTest emulator (details in §A.6).

### 6.1 Power Saving rApp

*Power-Saving rApp* [8] aims to reduce power consumption in RAN operations by intelligently deactivating or adjusting the number of active cells during periods of low demand [44], ultimately optimizing energy efficiency across the network. In our evaluation, we use an ML-based Power-Saving rApp that determines whether to activate, deactivate, or maintain the state of one or both collocated capacity cells in a typical cell site deployment scenario with one coverage cell and two capacity cells in each sector. The decision making in our Power-Saving rApp is based on the physical resource block (PRB) utilization data from all collocated coverage and capacity cells, including the serving cell and the neighboring cells. The model is implemented as a convolutional neural network (CNN) comprising one convolutional layer, one pooling layer, and two fully connected layers. This design is based on [13], which employs a 1D CNN followed by three dense layers. The model takes as input the PRB utilization of the serving sector's coverage cell and two capacity cells along with that of neighboring cells. The model produces six possible outputs: activating capacity cell 1, activating capacity cell 2, activating both capacity cells, deactivating capacity cell 1, deactivating capacity cell 2, or deactivating both capacity cells. Since the considered rApp is a multi-classification task, we use the targeted UAP in the evasion attack strategy.

## 6.2 Evaluation Metrics

We consider four metrics, including **Accuracy** and **APD** as mentioned in §5.2.1. Additionally, we evaluate the attack success rate (ASR) using two metrics: **Targeted Attack Success Rate (TASR)** and **Non-Targeted Attack Success Rate (NTASR)**.

$$\text{TASR} = \frac{N_{\text{successful targeted attacks}}}{N_{\text{total attempted attacks}}} \times 100\%, \quad \text{NTASR} = \frac{N_{\text{successful non-targeted attacks}}}{N_{\text{total attempted attacks}}} \times 100\% \quad (8)$$

where: $N_{\text{successful targeted attacks}}$ is the number of adversarial examples that were misclassified as the attacker's desired target class, and $N_{\text{total attempted attacks}}$ is the total number of adversarial attack attempts, and $N_{\text{successful non-targeted attacks}}$ is the number of adversarial examples that were misclassified, regardless of the target class Using these two metrics, we can assess both targeted and non-targeted attacks.

## 6.3 Results

We evaluate the efficacy of the proposed evasion attack (with Targeted UAP) strategy against the power saving rApp relative to the white-box approach on the "base" model using TASR, NTASR, and APD metrics. For this, we rely on a real-world city-scale mobile cellular network dataset for model training and (attack) testing. This dataset spans 40 days with 15-minute granularity and includes cell level PRB utilization data. Next, we assess the network performance impact of proposed attack on the power-saving rApp using the Keysight RICTest emulator [35] (details in §A.6).

*6.3.1 Analysis of Different Surrogate Models.* The cloning accuracies for the Base, MobileNet, ResNet, DenseNet, and 1L models are 0.9979, 0.9959, 0.9975, 0.9939, and 0.8450, respectively (under no attack at $\epsilon = 0$, not shown in Table 2). These results suggest that cloned models generally perform well compared to the base model. To explore the results further, we use the TASR, NTASR, and APD metrics. In Table 2, we observe that the APD for the Base model (Power-saving model) and other surrogate models increases as $\epsilon$ increases. This trend is consistent across all models, suggesting that the perturbations become more noticeable with larger $\epsilon$ values. TASR and NTASR generally increase with $\epsilon$, but not always. For example, in DenseNet, TASR decreased when $\epsilon$ increased from 0.1 to 0.2, likely due to the output being clipped to stay within the input range, which can prevent the attack from succeeding even with a higher $\epsilon$. MobileNet achieves the best TASR at $\epsilon = 0.5$ in black-box attacks, with acceptable performance compared to the white-box attack on the base model. It is also worth noting that the attack operates on two levels: targeted attacks aim for specific outputs, while non-targeted attacks degrade the model's performance by generating incorrect outputs.

*6.3.2 Analysis of PGMs.* To evaluate the effectiveness of Targeted UAPs (or TUPs, in short) compared to input-specific perturbations using methods, such as PGD and FGSM, we use $\epsilon = 0.5$ and generate perturbations based on 500 prediction samples from the rApp victim model. From Figure 6, the first plot (a) shows that PGD achieves the highest TASR when using

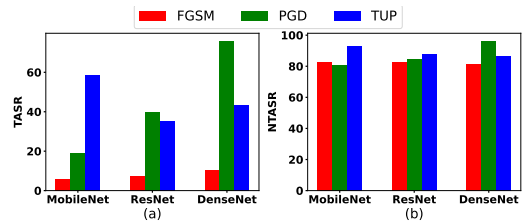

Fig. 6.  (a) TASR and (b) NTASR.

DenseNet, followed by TUP, which performs best with MobileNet. The second plot (b) indicates that the NTASR is high for all three methods, with PGD and TUP exhibiting similar performance. While PGD achieves a higher TASR, it has a higher APD of 1.402, compared to 1.171 for TUP at its highest TASR. Additionally, attacking 500 samples using DenseNet and PGD requires 29.75 minutes, whereas TUP, being a universal perturbation, is applied instantly.

This highlights PGD's scalability limitation—if a malicious rApp attempts to attack thousands of cells, the time required may be impractical. This difference in scalability highlights a key advantage of TUP: instead of generating a new perturbation for each input, the precomputed universal perturbation can be applied directly to any sample. As a result, TUP enables efficient large-scale attacks without the time overhead associated with input-specific methods. The potential scale of these attacks depends on the number of cells managed by the ML model; as the number of cells increases, the likelihood of larger-scale attacks also grow.

*6.3.3 Network Performance Results.* We use the Keysight RICTest emulator (as described in Section §A.6) to evaluate network performance under both normal and attack conditions. In Figure 7, the blue line represents the normal downlink (DL) throughput. This throughput varies in response to changes in the number of users distributed across the cells. Under normal conditions, both capacity and coverage cells are active, allowing users to be served by both types of cells. As a result, the DL throughput increases or decreases depending on the number of users being served.

In contrast, under the attack scenario, a sudden drop in DL throughput is observed, as shown by the red line in Figure 7 and marked by the vertical orange line. This decrease occurs because the attacker successfully deactivates two capacity cells during peak usage hours. As a result, all users from the deactivated capacity cell were shifted to the overlapping coverage cell. Consequently, the coverage cell had to serve both its existing users and the additional users from the turned-off capacity cells. This resulted in a degradation of the coverage cell's performance, as evidenced by the drop in throughput. Despite the attacker successfully turning off only two out of six capacity cells, we can observe a significant impact on the network throughput.

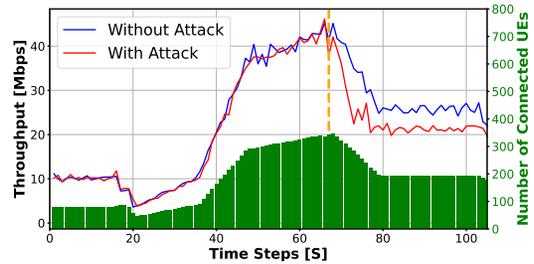

Fig. 7. Impact of successful attacks on capacity cells, causing reduced throughput.

## 7 Defense Evaluation

Our defense evaluation focuses on *adversarial training* (AT) [26] (training on perturbation-augmented data) and *defensive distillation* [27] (decision boundary smoothing via model distillation), as these are prominent defense techniques and provide practical robustness without compromising inference latency requirements[7].

From Figure 8 (a) for the IC xApp, our black-box attack strategy overcomes defensive distillation with only a 0.05 APD gap. We believe that this vulnerability stems from the effectiveness of our model cloning algorithm (§ 4.2.1), which replicates the victim model's decision boundaries despite distillation's theoretical protection. While distillation aims to reduce the gradient signals for attackers, we believe that the surrogate model's fidelity to the victim's behavior nullifies this advantage, enabling successful attacks within the cloned decision space. For our implementation of AT, we obtained the best defense when we generated adversarial variants of the 1,500 benign training samples across seven perturbation magnitudes ($\epsilon = [0.02, 0.05, 0.1, 0.2, 0.3, 0.4, 0.5]$), resulting in a final augmented training set of 10,500 adversarial samples combined with the original 1,500 benign samples. Here, we can see that AT outperforms distillation as a defense with an APD gap of 0.11, and our black box evasion attack strategy requires an APD gap of 0.16 to overcome this defense. Despite the relative improvement, the proposed strategy is still able to execute a successful attack with an APD of 0.35 being a hard boundary for defense. The analogous experiment for the power saving rApp over Non-RT RIC is shown in Figure 8 (b). Results are mostly consistent with the IC xApp. The base model is the most vulnerable, while adversarial training offers better protection than distillation defense in reducing TASR. However, distillation defense is more effective at lower epsilon values for TASR, although its performance varies at higher values. Note that adversarial training (AT) was implemented by augmenting the training set with adversarial examples generated using the same surrogate model (DenseNet121) employed by the attacker to simulate a realistic scenario.

---

[7]AML defenses fall into three categories, detection, data pre-processing, and robustness enhancement [73, 77]; the last is commonly instantiated via four tactics: gradient obfuscation [10], randomized smoothing [20], AT [26], and defensive distillation [27]. We adopt the latter two because there is no inference overhead.

We also believe that beyond these generic AML defenses, O-RAN-specific mitigation strategies could further enhance security. These may include runtime anomaly detection on Shared Data Layer (SDL) data streams, continuous behavioral monitoring of xApps and rApps post-deployment, and periodic audits of access control policies to prevent privilege escalation. Such mechanisms would help detect deviations from normal opera-tion and restrict malicious data manipulation, even in the presence of adversarial perturbations, and are promising avenues for future work.

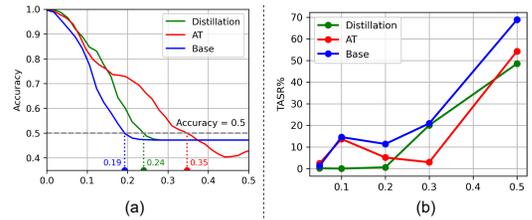

Fig. 8. Distillation AT performance: (a) Spectrogram IC xApp vs. APD, (b) rApp TASR vs. $\epsilon$.

## 8 Conclusions

We have aimed to explore data vulnerabilities of both xApps and rApps within the O-RAN architecture. We qualitatively analyzed the O-RAN security mechanisms and limitations specific to xApps and rApps and the data access mechanisms to inform our threat model. Taking into account the design considerations and timing constraints unique to O-RAN, we designed an effective black-box evasion attack strategy targeting ML-based Apps in near-RT and non-RT RICs in the O-RAN system, while leveraging established, proven adversarial ML techniques as its key building blocks. We evaluated the proposed attack strategy through a real-world LTE/5G O-RAN testbed for Near-RT RIC and Keysight emulator for Non-RT RIC, demonstrating the effectiveness of the proposed attack strategies, even in the presence of prominent defense techniques like defensive distillation and adversarial training. Future work will explore O-RAN-specific defense mechanisms such as runtime monitoring of SDL data integrity, behavioral attestation for xApps/rApps, and policy-based anomaly detection tailored to the RIC platforms. We also wish to further explore how perceptibility, particularly in the context of Average Perturbation Distance (APD), can be quantitatively analyzed to balance stealthiness and impact in near-real-time (near-RT) settings. In addition, it can be interesting to extend beyond classification-based use cases to include diverse O-RAN functions such as traffic steering, anomaly detection, and resource scheduling. This broader scope will enable assessing the applicability of our attack strategies across reinforcement learning and other advanced ML-driven control models.

## 9 Acknowledgments

The work of P. Gajjar, A. Ganiyu and V. K. Shah was supported by the Public Wireless Supply Chain Innovation Fund (PWSCIF) under Federal Award ID Number 51-60-IF007 and NSF award 2120411. The work of M. Khoja and M. Marina was supported in part by a project funded by the UK Department for Science, Innovation and Technology (DSIT). M. Marina's work was also supported by the UKRI/EPSRC grant UKRI860.

## A Appendix

### A.1 Related Work

---

**Algorithm 1:** Model Cloning Algorithm (MCA)

---

**Input:** List of classifier architectures $\mathcal{A}$, RIC database $\mathcal{D}_{RIC}$, xApps/rApps predictions $\mathcal{P}_{IC}$

**Output:** Cloned model $\mathcal{M}_C$ with best cloning accuracy

**Step 1: Creating the Cloning Dataset**

Collect cloning data $\mathcal{D}_{clone}$ by observing the victim xApp/rApp's predictions $\mathcal{P}_{IC}$ and the RIC data (Fig/KPMs) $\mathcal{D}_{RIC}$

**Step 2: Stratification**

Perform a stratified train-test split on $\mathcal{D}_{clone}$ to obtain training set $\mathcal{D}_{train}$ and validation set $\mathcal{D}_{val}$. Let $\mathcal{Y}_{clone}$ be the labels in $\mathcal{D}_{clone}$. The stratified split ensures:

$$\frac{|\mathcal{D}_{train}^c|}{|\mathcal{D}_{train}|} = \frac{|\mathcal{D}_{val}^c|}{|\mathcal{D}_{val}|} \quad \forall c \in \mathcal{Y}_{clone}$$

where $|\mathcal{D}_{train}^c|$ and $|\mathcal{D}_{val}^c|$ are the number of samples of class $c$ in the training and validation sets, respectively.

**Step 3: Train Classifier Architectures**

**for** *each classifier architecture* $\alpha \in \mathcal{A}$ **do**

  Initialize model parameters for $\alpha$;

  Set initial learning rate for $\alpha$;

  **while** *not converged* **do**

    Train $\alpha$ on $\mathcal{D}_{train}$;

    Compute validation loss $L_{val}(t)$ and validation accuracy on $\mathcal{D}_{val}$;

    **if** $L_{val}(t) - \min_{i \in [0,t]} L_{val}(i) > \delta$ *for k consecutive epochs* **then**

      Stop training (**Early Stopping**);

    **end**

    **if** $L_{val}(t) - \min_{i \in [0,t]} L_{val}(i) > \delta$ *for m consecutive epochs* **then**

      Update learning rate: $\eta_{new} = \eta_{old} \times \gamma$ (**Learning Rate Scheduler**);

    **end**

  **end**

  Save the best validation accuracy $acc_\alpha^{val}$;

**end**

**Step 4: Best Model Selection**

Select the model $\mathcal{M}_C$ with the highest validation accuracy:

$$\mathcal{M}_C = \arg\max_{\alpha \in \mathcal{A}} acc_\alpha^{val}$$

**Step 5: Return Best Cloned Model**

**return** $\mathcal{M}_C$

---

Security in mobile networks has been extensively studied, particularly in the context of 5G, Open RAN (O-RAN), and disaggregated network architectures. The study in [75] highlight critical vulnerabilities in the 5G fronthaul, demonstrating that the absence of mandatory integrity protection can be exploited to degrade network performance or launch large-scale denial-of-service (DoS) attacks. Similarly, the authors in [28] analyze O-RAN's open interfaces, emphasizing that disaggregation significantly increases the attack surface, necessitating stronger authentication and access control mechanisms. In addition, the study in [11] provides a comprehensive threat model of O-RAN vulnerabilities, identifying attack vectors against xApps, rApps, and external APIs. The §2.2.2 and §3 also corroborate these claims with an emphasis on NTIA's audit [48] and other resources like [33, 53, 74] that show how an authenticated xApp can be malicious and lead to adversarial attacks making AML a particularly relevant emerging threat.

Adversarial ML attacks have been widely studied in wireless networks. The authors in [40] demonstrated the practical vulnerabilities of ML-based wireless systems through the RAFA framework, showing that low-power adversarial perturbations can degrade ML-based communication and sensing systems. Other works have explored black-box adversarial attacks in spectrum sharing environments [63], RL-based spectrum access [72], and deep learning-based wireless communications [62]. Another work, [70] focuses on the cellular physical layer and apply adversarial attack on three different ML models using different training methods: supervised, unsupervised and reinforcement learning.

These studies establish that ML-based wireless applications are highly vulnerable to adversarial perturbations, raising concerns for AI-driven RAN optimizations in O-RAN.

While adversarial threats have been explored in O-RAN AI-driven control, prior research has primarily focused on white-box attacks, where the adversary has full knowledge of the model. For example, the authors in [21] employed FGSM-based adversarial policy infiltration to manipulate RAN control decisions. Their study demonstrated that spoofed signal power measurements and altered channel parameters could significantly degrade user data rates and impair ultra-reliable low-latency communication (URLLC) services. Another study [22] investigated adversarial vulnerabilities in near-real-time vehicular networks, where small perturbations in environmental state data led to severe performance degradation in resource allocation and increased packet loss. However, both studies assume full model access for the attacker, making them less applicable to real-world O-RAN deployments where black-box attacks are more likely.

Unlike prior works that focus on white-box attacks or theoretical AML threats, our work presents the first real-world analysis of black-box evasion attacks on AI-driven xApps and rApps in O-RAN deployments, leveraging a real-world LTE/5G O-RAN testbed (with OSC RIC) for Near-RT RIC attacks and Keysight RICtest emulation platform for Non-RT RIC attacks. Unlike prior studies that focus on gradient-based attacks with full model access, we demonstrate that transferability based black-box evasion attacks leveraging inference data access and model cloning are both feasible and highly effective, even when adversaries have no knowledge of the target models. Our results highlight security gaps in AI-driven RAN control and emphasize the need for more focus on robust adversarial ML defense mechanisms to clear the way for widespread real-world O-RAN deployments.

## A.2 Further Details on Attack Design

This section contains the MCA Algorithm 1, and UAP generation Algorithm 2.

## A.3 Perturbation Generation Methods

As shown in §4.2.2, our work leverages two types of perturbation generation methods (PGMs), namely, Norm-Bounded Methods and Norm-Unbounded Methods.

***Norm-Bounded Methods.*** These generate perturbations constrained by a specific norm (e.g., $L_2$ or $L_\infty$ norm) to ensure the perturbations are imperceptible or minimally perceptible to the human eye. The following are some common norm-bounded methods:
• **Fast Gradient Sign Method (FGSM) [26]**: FGSM is a simple and efficient method for generating adversarial examples. It works by computing the gradient of the loss for the input data and then perturbing the sample data in the direction of the gradient. The perturbation is scaled by a small factor $\epsilon$ to ensure it stays within a specified norm bound. The adversarial example $\mathbf{x}'$ is computed as:

$$\mathbf{x}' = \mathbf{x} + \epsilon \cdot \text{sign}(\nabla_\mathbf{x} J(\theta, \mathbf{x}, y)) \tag{9}$$

where $J(\theta, \mathbf{x}, y)$ is the loss function, $\theta$ represents the model parameters, and $y$ is the true label.
• **Projected Gradient Descent (PGD) [43]**: PGD is an iterative version of FGSM that applies the perturbation multiple times with smaller steps. After each step, the perturbed sample is projected back onto the feasible set defined by the norm bound. The iterative process is defined as:

$$\mathbf{x}_{t+1} = \text{Proj}_{\mathbf{x}+\epsilon}(\mathbf{x}_t + \alpha \cdot \text{sign}(\nabla_\mathbf{x} J(\theta, \mathbf{x}_t, y))) \tag{10}$$

where $\alpha$ is the step size and Proj denotes the projection operator that ensures the perturbation remains within $(-\epsilon, \epsilon)$.

***Norm-Unbounded Methods.*** Norm-unbounded methods do not constrain the perturbation within a specific norm, allowing for potentially larger and more noticeable perturbations. Some prominent norm-unbounded methods include:
• **Carlini & Wagner (C&W) [15]**: The C&W attack is an optimization-based method that generates adversarial examples by minimizing the perturbation while ensuring the perturbed sample is misclassified. The attack solves the following optimization problem:

$$\min_\mathbf{r} \|\mathbf{r}\|_2 + c \cdot f(\mathbf{x} + \mathbf{r}) \tag{11}$$

where $f(\mathbf{x} + \mathbf{r})$ is a function that measures the confidence of the classifier in the original class, and $c$ is a constant that balances the perturbation magnitude and the misclassification objective. The optimization is typically solved using gradient descent [15].
• **DeepFool [47]**: DeepFool is an iterative attack that finds the smallest perturbation required to change the classification of an input sample. It works by approximating the classifier's decision boundary as a hyperplane and iteratively perturbing the input to cross this boundary. At each iteration, the perturbation $\mathbf{r}_i$ is computed as:

$$\mathbf{r}_i = \frac{f(\mathbf{x}_i)}{\|\nabla f(\mathbf{x}_i)\|_2^2} \nabla f(\mathbf{x}_i) \tag{12}$$

where $f(\mathbf{x})$ is the output of the classifier and $\nabla f(\mathbf{x})$ is the gradient of the classifier's output for the input.

## A.4 LTE/5G O-RAN Testbed

Similar to [18], we prototype an over-the-air O-RAN testbed that comprises a Core, RAN, UE, a jammer, and the near-RT RIC (See Fig. 9 (I)). The RAN/Core and the UE are based on the open-source srsRAN cellular software stack, designed for building LTE/5G cellular networks. Both components operate on Ubuntu 20.04 OS and run on Intel Core i7-9700 processors with 8 CPU cores, 32GB RAM, 8 threads, and a clock speed of 3.0GHz. We modified the srsRAN codebase to include a buffer for storing I/Q samples and added RAN control capabilities, switching between adaptive or fixed MCS. The RAN and UE use USRP B210s for RF front-end operations. The near-RT RIC is hosted on a rack server with an AMD EPYC$^{\text{TM}}$ 7443P with 24 CPU cores, 48 threads, 64GB RAM, and a base clock of 2.85GHz. The near-RT RIC implemented with the OSC open-source codebase [54] is compiled on the server. It interfaces with the RAN via an E2-lite[8] interface, enabling real-time decision-making based on network conditions. The jammer generates signals that interfere with the UE's uplink signal, transmitted over-the-air (OTA). The interference signals are essential for testing the resilience of the network to adversarial interference; these are generated by the jammer and transmitted using GNURadio.

---

**Algorithm 2:** Universal Adversarial Perturbation (UAP) Generation

**Input:** Samples $S$, classifier $C$, radius $\epsilon$, target fooling $\zeta$
**Output:** Universal vector $\mathbf{u}$
Initialize $\mathbf{u} \leftarrow 0$;
**while** $Error(S, \mathbf{u}) \leq 1 - \zeta$ **do**
  **for** $\mathbf{x}_i \in S$ **do**
    **if** $C(\mathbf{x}_i + \mathbf{u}) = C(\mathbf{x}_i)$ **then**
      Compute minimal $\Delta\mathbf{u}_i$ s.t.
      decision flips;
      $\mathbf{u} \leftarrow \mathcal{P}_{p,\epsilon}(\mathbf{u} + \Delta\mathbf{u}_i)$;
    **end**
  **end**
**end**
**return** $\mathbf{u}$;

---

## A.5 Data Generation Process for Evaluating Spectrogram/KPM-based IC xApp

The *spectrogram dataset* consists of 3,000 spectrograms equally divided into two classes: 1,500 representing uplink UE signals of interest (SOI) without interference, and 1,500 containing continuous wave interference (CWI) combined with the SOI. Each spectrogram is an RGB image of size (128, 128, 3), generated from OTA LTE uplink transmissions at 2.56 GHz with 25 PRBs (5 MHz bandwidth), sampled at 7.68 MSps using the srsRAN 4G stack. TCP traffic at 5 MHz was transmitted between the UE and RAN via iperf3. The CWI signals, transmitted at the same uplink frequency as the SOI, were generated using GNURadio and USRP with gain values ranging from 40 dB to 45 dB. To construct the $\mathcal{D}_{clone}$ database as mentioned in §4.2.1 for attacking the spectrogram-based victim xApp, we observe the victim xApp's predictions $\mathcal{P}_{IC}$ and dataset $\mathcal{D}_{RIC}$ for 3,000 spectrograms. A sample perturbed image that successfully attacks the victim xApp and the computed UAP are shown in Figure 9 (II). This is computed by using FGSM and a MobileNetV2-based surrogate model with an $\epsilon$ value of 0.05. The *KPM dataset* consists of 2,910 KPM instances, equally distributed between interference and no interference scenarios. The same experimental setup was used to collect uplink metrics, specifically SINR, bitrate/throughput, BLER, and MCS, sampled at intervals of 0.5 to 1 second. The $\mathcal{D}_{clone}$ database for attacking the KPM-based xApp is obtained by observing $\mathcal{P}_{IC}$ and $\mathcal{D}_{RIC}$ for 1,455 KPM instances.

---

[8]We adopt a lightweight E2-lite interface developed in [18] that facilitates closed-loop communication between the near-RT RIC and the RAN, and utilizes SCTP protocol as the underlying transport protocol.

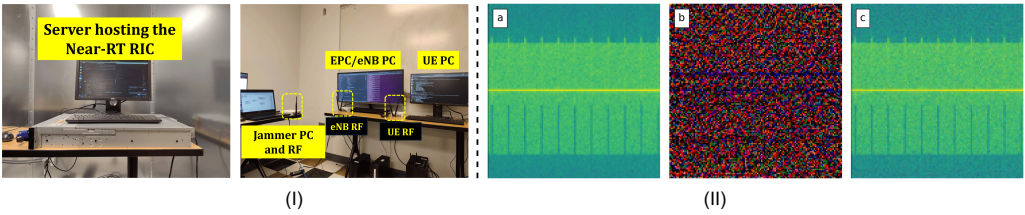

Fig. 9. (I) LTE/5G O-RAN testbed; (II) spectrogram: (a) original, (b) UAP, (c) perturbed.

## A.6 Keysight RICTest Emulator Details

To validate the targeted evasion attack on the power-saving rApp's ML model, the Keysight RICtest emulator is used [35]. RICtest is a testing tool designed for RAN Intelligent Controllers. It simulates the behavior of UE in terms of data and mobility, as well as models the operation of hundreds of cells. The tool provides near-real-time telemetry and stateful control for O-RAN nodes, enabling closed-loop validation. RICtest supports O-RAN standards-based protocols, such as the E2, O1, and A1 interfaces. In our specific use case, RICtest is utilized to implement a Power Savings ML model (via the O1 interface) with offline training. This ML model helps control the activation and deactivation of capacity cells to reduce power consumption, ensuring that user experience is not compromised. The RAN setup (Figure 10) comprises one O-gNB with three coverage cells and six capacity cells. The coverage cells operate on frequency band 77 and provide coverage over an area of approx. 2 km. The capacity cells operate on frequency band 79, with each capacity cell covering an area of about 0.3 km. Each coverage cell overlaps with two capacity cells.

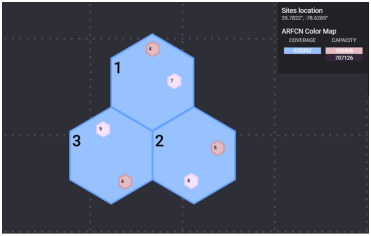

Fig. 10. The Keysight emulator has three primary coverage cells (1, 2, and 3) with each containing two smaller capacity cells: Cell 1 (4, 7), Cell 2 (5, 8), and Cell 3 (6, 9). Different colors indicate different frequencies.

In this setup, the UEs are distributed between the coverage and capacity cells, with the coverage cell having lower priority than the capacity cell. Low number of 10 UEs are connected to each coverage cell, while a dynamic number of UEs are connected to the capacity cells. The number of UEs in the capacity cell varies, ranging from 0 to 55, increasing and decreasing over time. The UE traffic primarily consists of downlink traffic, with a mix of steady and bell-curve traffic patterns. The primary KPIs for user performance and RAN load are downlink throughput, and "RRU.PrbTotDl," which reflects the total percentage usage of physical resource blocks on the downlink, serving as a measure of the load for each cell. Additionally, "RRC.ConnMean" represents the mean number of users in RRC connected mode. The ML model is trained offline using RRU.PrbTotDl KPI data collected from both capacity and coverage cells over the past 12 timesteps. Using this data, the model determines when to deactivate and reactivate the capacity cells. At the beginning, the power-saving rApp sends a data collection request through the SMO, which then forwards it through the O1 interface to the gNB. This request collects the emulated RAN network configuration, including the coverage cell, capacity cell, and the current status of each of these cells. Based on the PRB input, the rApp decides whether or not to turn off the capacity cell. If the decision is made to turn off the cell, a command is sent through the SMO via the O1 interface to the O-gNB to update the status of the capacity cell. A feedback response is then sent back to the rApp, confirming that the action was completed.